# Angular Scattering in Charge Exchange: Issues and Implications for Secondary Interstellar Hydrogen

P. Swaczyna[1,*], D. J. McComas[1], E. J. Zirnstein[1], and J. Heerikhuisen[2]

[1]Department of Astrophysical Sciences, Princeton University, Princeton, NJ 08544, USA
[2]Department of Mathematics and Statistics, University of Waikato, Hamilton 3240, New Zealand

**Abstract**

Interstellar neutral atoms provide a remote diagnostic of plasma in the outer heliosheath and the very local interstellar medium via charge exchange collisions that convert ions into atoms and vice versa. So far, most studies of interstellar atoms assumed that daughter hydrogen atoms directly inherit the kinetic properties of parent protons. This assumption neglects angular scattering of the interacting particles. However, for low relative velocities, as expected for charge exchanges in the outer heliosheath, this scattering is significant. In this study, we present how the parameters of daughter populations depend on the relative velocity and temperatures of parent populations. For this purpose, we numerically compute collision terms with and without this scattering. We find that the secondary population of interstellar hydrogen atoms, for the parent populations with the relative bulk velocity of 20 km s$^{-1}$ and equal temperatures of 7500 K, has ~2 km s$^{-1}$ higher bulk velocity if the scattering is taken into account. Additionally, temperatures are higher by ~2400 K and ~1200 K in parallel and perpendicular direction to the relative motion of parent populations, respectively. Moreover, a significant departure of secondary atoms from the Maxwell-Boltzmann distribution is expected for high relative velocities of parent populations. This process affects the distribution and density of interstellar atoms in the heliosphere and production of pickup ions. Thus, we show that angular scattering in charge exchange collisions is important to include in analyses of interstellar neutral atoms and pickup ions observed at 1 au and in the outer heliosphere.

1. ## Introduction

The heliosphere, created by the interaction of the expanding solar wind and the very local interstellar medium (VLISM), leads to separation of VLISM neutral and ionized components (Parker 1961; Blum & Fahr 1970). Interstellar neutral (ISN) atoms, in contrast to charged particles, freely penetrate the heliosphere. Atoms and ions are coupled by charge exchange collisions, which continuously mix these components, especially hydrogen atoms and protons. As a result, ISN atoms inside the heliosphere need to be described by at least two populations (Baranov & Malama 1993; Holzer 1977; Wallis 1975). The primary population originates far outside the heliopause, where plasma and neutrals flow together in the pristine interstellar medium. The secondary population is created by charge exchange in the outer heliosheath, surrounding the heliopause, where the properties of neutrals and plasma differ.

The relative abundance of secondary and primary ISN atoms depends on the charge exchange rate in the outer heliosheath. Observations of ISN hydrogen atoms using the Ly$\alpha$ line (Lallement & Bertaux 1990; Lallement et al. 1993; Quémerais et al. 1999; Lallement et al. 2005) show that the average inflow direction of ISN hydrogen atoms significantly differs from the inflow of ISN helium atoms (Witte et al. 2004; Bzowski et al. 2014, 2015; McComas et al. 2015; Schwadron et al. 2015; Wood et al. 2015; Swaczyna et al. 2018). This discrepancy is a result of the secondary population that cannot be distinguished from the primary population in Lyα observations.

---

[*] Corresponding author: swaczyna@princeton.edu

ISN atoms are also observed directly inside the heliosphere. ISN helium atoms were directly sampled by the GAS instrument on Ulysses (Witte et al. 1996). Currently, IBEX-Lo (Fuselier et al. 2009; Möbius et al. 2009b) on the *Interstellar Boundary Explorer* (McComas et al. 2009) provides observations of ISN hydrogen (Möbius et al. 2009a; Saul et al. 2012; Galli et al. 2019), helium (Bzowski et al. 2012; Möbius et al. 2012; McComas et al. 2015), oxygen, and neon (Park et al. 2014, 2015). Direct observations of ISN helium atoms show two components related to the primary and secondary populations (Kubiak et al. 2014, 2016; Bzowski et al. 2017; Wood et al. 2017). Moreover, the distribution of ISN helium atoms may depart from the thermal equilibrium (Swaczyna et al. 2019; Wood et al. 2019).

Interpretation of ISN hydrogen data is more complicated because of a significantly higher number of secondary atoms and due to ionization losses into 1 au and radiation pressure, which modifies the trajectories of hydrogen atoms (Axford 1972). These effects make interpretation of hydrogen observations more complex (Schwadron et al. 2013; Katushkina et al. 2015). Recently, Kowalska-Leszczyńska et al. (2018a, 2018b) developed a new model of the solar Lyα line, which is necessary to trace trajectories of ISN hydrogen atoms in the heliosphere.

So far, heliospheric studies have assumed that charge exchange between hydrogen atoms and protons does not lead to angular scattering of velocities of the interacting particles. Under this assumption, this process can be fully described by the energy-dependent charge exchange cross section. In most cases, compilations of experimental data are used via analytic formulae (e.g., Barnett 1990; Lindsay & Stebbings 2005). In reality, the velocities of colliding particles also change directions, with the scattering angle between incoming and outgoing particle velocity vectors. In this case, the charge exchange needs to be described by a differential cross section that depends on the scattering angle, typically defined in the center-of-mass (CM) frame. Izmodenov et al. (2000) noted that the elastic scatterings and charge exchange between hydrogen atoms and protons could be generally described as one process between indistinguishable particles. Heerikhuisen et al. (2009) checked the consequences of angular scattering in charge exchange collisions for the global modeling of the heliosphere and concluded that this scattering does not significantly change the global distribution of plasma.

Differential cross sections for the charge exchange between hydrogen atoms and protons is mostly known from theoretical calculations (Hodges & Breig 1991; Krstic & Schultz 1999; Schultz et al. 2008). Recently, Schultz et al. (2016) provided a comprehensive set of cross sections for energies from $10^{-4}$ eV to $10^{4}$ eV. In this study, we apply the results of Schultz et al. (2016) to determine the impact of the momentum transfer due to scattering in charge exchange collisions on distributions of the secondary populations of ISN hydrogen atoms.

**2. Hydrogen-Proton Charge Exchange**
Charge exchange cross sections depend on the collision energy that expressed as the projectile energy ($E_{\text{proj}}$), in which one of the colliding particles is at rest, or as the center-of-mass (CM) energy ($E_{\text{CM}}$). Alternatively, the cross section can be expressed as a function of the relative speed. The projectile energy is most useful when one of the colliding particles is much more energetic in the used reference frame.

However, in this study, the particles have comparable speeds in a heliospheric frame, and therefore, we use the CM energy. Moreover, scattering angles for differential cross sections are also defined in the CM frame. For conversion from studies that use the projectile energy, we use the approximate formula $E_{CM} = E_{proj}/2$, neglecting the small difference in mass between hydrogen atoms and protons.

Collisions between protons and hydrogen atoms can result in electron transfer from atom to proton or in elastic scattering of interacting particles. At low energies, charge transfers are mostly through the resonant channel, i.e., outgoing atoms are in the same energy state as incoming atoms and, hence, energy and momentum are conserved. Therefore, charge exchange and elastic collisions between hydrogen atoms and protons cannot be distinguished because a charge exchange collision in angle $\theta$ and an elastic collision in angle $180° - \theta$ give the same final state. This quantum indistinguishability is overcome if spins of colliding particles are known. Schultz et al. (2016) showed that the difference between the sum of the integral cross sections of elastic scattering and charge exchange for particles with known spins and the elastic scattering approach for indistinguishable particles is smaller than 5% for CM energies larger than 0.01 eV and significantly decreases for higher energies. Therefore, for heliospheric studies, separation of elastic and charge exchange collisions is justified.

*2.1. Integral Cross Sections*

In this subsection, we compare the integral cross sections from Schultz et al. (2016) with the cross sections widely used in the heliospheric community. The most commonly used $H^+$–$H^0$ charge exchange cross section is given in Lindsay & Stebbings (2005). They provided an analytic formula that applies for the CM energy range 2.5 eV – 125 keV. Another formula is given in Barnett (1990) for the CM energy range 0.12 eV – 630 keV. Both of these sources combined pre-existing experimental measurements; however, Barnett (1990) additionally used theoretical calculations to extend to lower energies.

Figure 1 presents a comparison of the cross sections calculated from the formulae given by Barnett (1990) and Lindsay & Stebbings (2005) with the numerical charge exchange cross section from Schultz et al. (2016). The experimental measurements start at 2.5 eV (Belyaev et al. 1967), which therefore is the lower limit of the recommended formula by Lindsay & Stebbings (2005). There is good agreement between these cross sections for the well-measured range 100 – 1000 eV. However, the cross sections tend to diverge for lower energies with approximately ~20% variation around the results of Schultz et al. (2016). A structure visible around ~4 keV in the cross section from Schultz et al. (2016) may be a result of low accuracy of their calculations for energies above 1 keV. Note that collision energies for charge exchange between the ISN atoms and outer heliosheath plasma are much smaller than 1 keV, and therefore, these discrepancies do not affect the presented study.

Many heliospheric studies use the cross section from Lindsay & Stebbings (2005), even outside of the energy range recommended by authors. The comparison presented in Figure 1 shows that this is especially problematic for lower energies that are necessary for collisions in the outer heliosheath. We propose an update based on results from Schultz et al. (2016) for CM energies from 0.01 eV to 100 eV, and the formula from Lindsay & Stebbings (2005) for CM energies from 100 eV to 125 keV. The updated cross section can be calculated for projectile energy $E_{proj}$ in keV from equation

$$\sigma_{cx}(E_{proj}) = \left(4.049 - 0.447 \ln E_{proj}\right)^2 \left[1 - \exp\left(-\frac{60.5}{E_{proj}}\right)\right]^{4.5} \times 10^{-16} \text{ cm}^2. \quad (1)$$

The values obtained from the above equation differ less than 10% from the Lindsay & Stebbings (2005) for CM energies higher than 100 eV, i.e., comparable with the accuracy of their formula. At the same time, this formula extends the lower energy range to 0.01 eV, i.e., typical collision energies in the outer heliosheath.

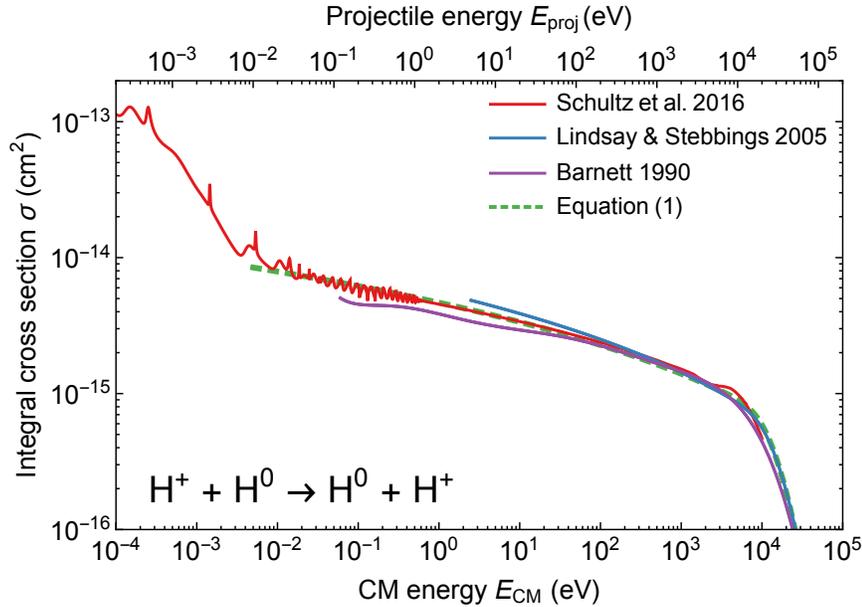

**Figure 1.** Comparison of the integral $H^+$–$H^0$ charge exchange cross sections widely used in other heliospheric studies (Barnett 1990; Lindsay & Stebbings 2005) with the one used here (Schultz et al. 2016). We also present the analytical approximation given in Equation (1), see text for details.

*2.2. Differential cross section*

Two angles can describe the collisional scattering of particles: the scattering angle between the initial and final velocities in the CM frame, and the azimuthal angle around the incoming particle velocity (see Figure 2). Due to symmetry, differential cross sections for binary collisions depend only on the first one. Therefore, differential cross sections can be presented as integrals over the azimuthal angle: $\frac{d\sigma}{d\Omega} 2\pi \sin(\theta)$, where $\theta$ is the CM scattering angle.

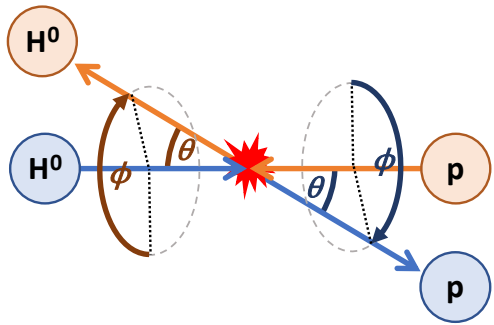

**Figure 2**. Schematic illustration of a charge exchange collision in the CM frame. The direction of the outgoing particles is described by the angle between the inintial and final velocity $\theta$ and the azimuthal angle $\phi$. Selection of the starting point of the azimuthal angle is arbitrary.

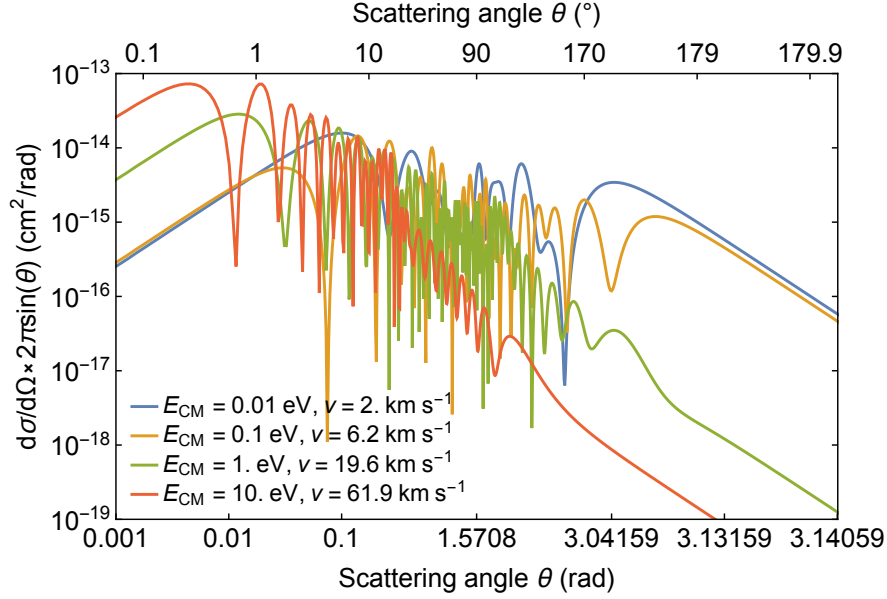

**Figure 3.** Differential cross section for the charge exchange in $H^+$–$H^0$ collisions for $E_{CM} = 0.01, 0.1, 1, 10$ eV.

Figure 3 presents the differential cross sections from Schultz et al. (2016) for the charge exchange in $H^+$–$H^0$ collisions for select CM energies in the range considered in this study. The character of the scattering significantly changes in this energy range. At lower energies, the distribution of scattering angles is broad, and therefore, the momentum exchange between nuclei may be significant. As the collision energy grows, scattering angles decrease, and thus the relative change of momenta of colliding particles is smaller. The differential cross sections show oscillations resulting from summation over the partial wave contributions. In this study, the scattering angle, i.e., the angle each incoming nucleus scatters, is defined between the incoming proton (or hydrogen atom) and the outgoing hydrogen atom (proton). Figure 4 shows the mean and median scattering angle in charge exchange collisions as a function of energy and relative speed. The mean angle is slowly changing up to $E_{CM} \approx 1$ eV. Above this energy, it drops significantly from ~45° to ~2° at 100 eV. The median angle is smaller than the mean angle, therefore, scattering angles smaller than the mean scattering angle are more frequent than the angles above this value. This figure clearly shows that negligible scattering angles are not justified in charge exchange collisions between plasma ions and neutrals in the outer heliosheath since bulk flow speeds are typically less than ~25 km s$^{-1}$.

## 3. Production Rate of Secondary ISN Atoms

Production of the secondary population in the outer heliosheath can be described by a collisional term in the Boltzmann transport equation. In this section, we present a commonly used form for this source term under the no scattering approximation (Section 3.1) as well as the framework for the production rate including angular scattering in charge exchange collisions (Section 3.2). The production rates depend on the properties of proton and neutral atom distributions: $f_p(\boldsymbol{x}, \boldsymbol{v}, t)$ and $f_H(\boldsymbol{x}, \boldsymbol{v}, t)$, which vary in space and time. This variation is important when the Boltzmann equation is solved, but we omit this dependency in the presented formulation because here we consider local and time-independent production.

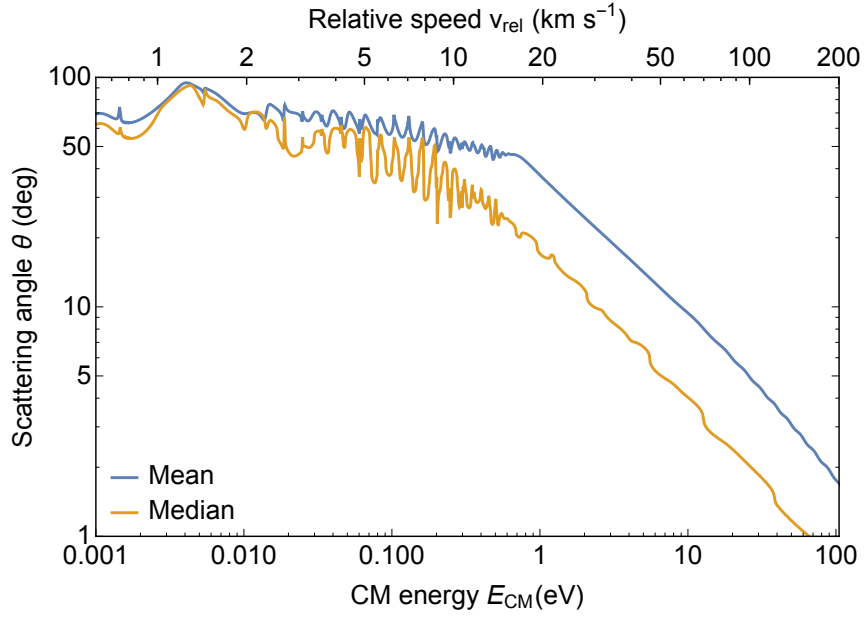

**Figure 4.** Mean and median CM scattering angle in $H^+$–$H^0$ charge exchange collisions as a function of the CM energy (bottom scale) and the relative speed (top scale). Median values smaller than mean values show that scattering angles smaller than the mean scattering angle are more probable.

### 3.1. Without Angular Scattering

Most outer heliosheath studies neglect angular scattering in charge exchange collisions and assume that the production rate of secondary neutral atoms is directly proportional to the VDF of the parent protons $f_p(\boldsymbol{v})$ (e.g., Ripken & Fahr 1983; Pauls et al. 1995; Izmodenov et al. 2001; Zank et al. 2009):

$$\left.\frac{\partial f_H(\boldsymbol{v})}{\partial t}\right|_{P,NAS} = \beta_{cx}(\boldsymbol{v}) f_p(\boldsymbol{v}), \tag{2}$$

where $\beta_{cx}(\boldsymbol{v})$ is the charge exchange reaction rate. This rate can be fully expressed by an integral over the VDF of neutral hydrogen $f_H(\boldsymbol{v}_H)$:

$$\beta_{cx}(\boldsymbol{v}) = \int |\boldsymbol{v} - \boldsymbol{v}_H| \sigma_{cx}(|\boldsymbol{v} - \boldsymbol{v}_H|) f_H(\boldsymbol{v}_H) d^3\boldsymbol{v}_H. \tag{3}$$

In this equation, $\sigma_{cx}(|\boldsymbol{v} - \boldsymbol{v}_H|)$ denotes the integral charge exchange cross section. Note that the relative velocities $|\boldsymbol{v} - \boldsymbol{v}_H|$ are sufficient to calculate CM and projectile energies, and therefore can be used to represent the dependence of the cross section with energy. It is often assumed that this cross section does not significantly vary over the velocity range of hydrogen atoms, and therefore the rate can be approximated as follows

$$\beta_{cx} = v_{rel} \sigma_{cx}(v_{rel}) n_H, \tag{4}$$

where $v_{rel}$ denotes the mean relative speed, and $n_H$ is the density of neutral hydrogen. For neutral hydrogen atoms with a bulk speed $\boldsymbol{u}_H$ and temperature $T_H$, the mean relative speed can be calculated analytically (Ripken & Fahr 1983; Pauls et al. 1995)

$$v_{\text{rel}} = u_{\text{T,H}} \left[ \left( \omega + \frac{1}{2\omega} \right) \text{erf}(\omega) + \frac{\exp(-\omega^2)}{\sqrt{\pi}} \right], \quad (5)$$

where $u_{\text{T,H}} = \sqrt{2k_B T_H/m_H}$ is the thermal velocity, and $\omega = |\boldsymbol{v} - \boldsymbol{u}_H|/u_{\text{T,H}}$ is the ratio of the relative bulk velocity to the thermal speed.

*3.2. With Angular Scattering*

To account for angular scattering of velocities, a more general form of the collision production term is

$$\left.\frac{\partial f_H(\boldsymbol{v})}{\partial t}\right|_{\text{P,WAS}} = \int |\boldsymbol{v}_p - \boldsymbol{v}_H| \frac{d\sigma_{\text{cx}}}{d\Omega}\left(|\boldsymbol{v}_p - \boldsymbol{v}_H|, \theta\right) f_H(\boldsymbol{v}_H) f_p(\boldsymbol{v}_p) \times$$

$$\delta^{(3)}\left(\boldsymbol{v} - \boldsymbol{v}'(\boldsymbol{v}_p, \boldsymbol{v}_H, \theta, \phi)\right) d^3\boldsymbol{v}_H d^3\boldsymbol{v}_p \sin\theta \, d\theta d\phi, \quad (6)$$

where the integral is over the full velocity space of the VDFs of hydrogen atoms and protons, scattering angle $\theta$, and azimuthal angle $\phi$. Only some combinations of these quantities result in the production of atoms with velocity $\boldsymbol{v}$. Hence, the three-dimensional Dirac delta function $\delta^{(3)}$ selects a post-collision velocity $\boldsymbol{v}'(\boldsymbol{v}_p, \boldsymbol{v}_H, \theta, \phi)$ equal to the velocity of interest $\boldsymbol{v}$. This delta function allows to integrate out one of the velocities, which gives the form of the production term that is widely used (see e.g., Chapman & Cowling 1970). The integral is performed in any reference frame; however, the differential cross section is defined in the CM frame. Therefore, in the calculation of the post-collision velocity of the newly created hydrogen atom we account for the appropriate reference frame

$$\boldsymbol{v}'(\boldsymbol{v}_p, \boldsymbol{v}_H, \theta, \phi) = \frac{m_p}{m_H} \boldsymbol{r}(\boldsymbol{v}_p - \boldsymbol{v}_{\text{CM}}, \theta, \phi) + \boldsymbol{v}_{\text{CM}}, \quad (7)$$

$$\boldsymbol{v}_{\text{CM}}(\boldsymbol{v}_p, \boldsymbol{v}_H) = \frac{(m_p \boldsymbol{v}_p + m_H \boldsymbol{v}_H)}{m_p + m_H}. \quad (8)$$

In the CM reference frame, a newly created hydrogen atom conserves the absolute value of momentum of the incoming proton, but the velocity direction is rotated $\boldsymbol{r}(\boldsymbol{v}_p - \boldsymbol{v}_{\text{CM}}, \theta, \phi)$. The rotation around the azimuthal angle $\phi$ is arbitrary due to symmetry around the relative velocity of colliding particles (see Figure 2). The form of the rotation used for this study is presented in Appendix A.

We note that the general form of the production rate given in Equation (6) leads to the approximation given in Equations (2) and (3) after assuming that the differential cross section has a Dirac delta form around $\theta = 0$ proportional to the integral cross section, and that the masses of protons and hydrogen atoms are equal $m_p = m_H$. In this situation, the integral over angles is trivial, and Equation (7) gives that $\boldsymbol{v}' = \boldsymbol{v}_p$. Consequently, the integral over the incoming proton velocity is also trivial due to the Dirac delta function in Equation (6).

## 4. Distribution of Secondary Atoms

The production rates of secondary neutral atoms presented in Sections 3.1 and 3.2 are the collision terms in the Boltzmann transport equation. Solving the Boltzmann equation requires accounting for losses and space and time variation of plasma condition in the outer heliosheath. In this study, we compare the properties of these production rate distributions solely, even though they do not represent the actual population of secondary atoms. However, these findings can be easily applied in other heliospheric studies.

At the same time, the differences between the production rates in the two discussed approach give insight into the role of angular scattering in charge exchange collisions.

In this study, we consider the charge exchange process between protons and hydrogen atoms, which both follow Maxwell-Boltzmann distributions. This assumption is an idealized approach because the distributions of protons and neutral hydrogen atoms in the heliospheric studies are often non-Maxwellian. Nevertheless, this assumption gives insight into the role of this scattering. Temperatures and bulk velocities of these two populations are not necessarily equal. Without loss of generality of the problem, we use the frame of reference co-moving with the parent protons and assume the bulk velocity of hydrogen atoms is aligned with the Z axis. In these coordinates, we numerically compute the production rates in Equation (6) using the Monte Carlo method discussed in Appendix B. For the simplified approach without momentum transfer, we numerically integrated Equation (3) to account for the variation of the charge exchange cross section with relative speed.

The grid of parameters for the parent populations covers the possible range of these parameters in the outer heliosheath. Namely, the relative bulk velocities are taken from 0 km s$^{-1}$ to 50 km s$^{-1}$, every 5 km s$^{-1}$, and temperatures of protons and hydrogen atoms are 7500 K, 15000 K, and 22500 K. These temperatures correspond to the thermal velocities (calculated as $\sqrt{2k_\text{B}T/m_\text{H}}$ ) of 11.1 km s$^{-1}$, 15.7 km s$^{-1}$, and 19.3 km s$^{-1}$, respectively. In general, the relative velocity between plasma and neutrals should not exceed the ISN inflow velocity of ~25.4 km s$^{-1}$, as measured by IBEX (McComas et al. 2015). The temperatures are selected to cover the plasma temperature in the outer heliosheath (Zank et al. 2013; Heerikhuisen et al. 2016).

*4.1 Mean Velocity and Temperature of the Production Rate Distribution*
We characterize the calculated production rate of secondary atoms using distribution moments. The mean values in each of the three dimensions give bulk velocities, which are aligned with the relative motion of the parent population. The second central moments are calculated with the bulk velocity subtracted and are expressed in this study in Kelvin to facilitate comparison with the parental populations. They are not physical temperatures of the produced populations because (1) they are obtained from the production rate and not the actual VDF and (2) the population of secondary atoms is not following the Maxwell distribution function. Nevertheless, we call them "temperatures" for simplicity, and they should read as a measure of the mean energy of particles. We determine them separately in directions parallel and perpendicular to the relative velocity of the parent populations.

Figure 5 presents the bulk velocities for the two approaches and the difference between them as a function of the relative bulk velocity for select temperatures of the parent populations. Interestingly, the largest bulk velocities in the plasma frame are obtained in the approach without angular scattering and they are directed oppositely to the incoming neutral hydrogen atoms. This is caused by the charge exchange rate given in Equation (3), which is proportional to the relative velocity, and thus protons moving towards the neutral population are preferentially charge-exchanged. This effect is stronger for higher temperatures of protons and reduced for higher temperatures of neutral hydrogen atoms. The largest bulk velocity is expected for the relative neutral-plasma velocity slightly larger than the plasma thermal speed for each considered plasma temperature.

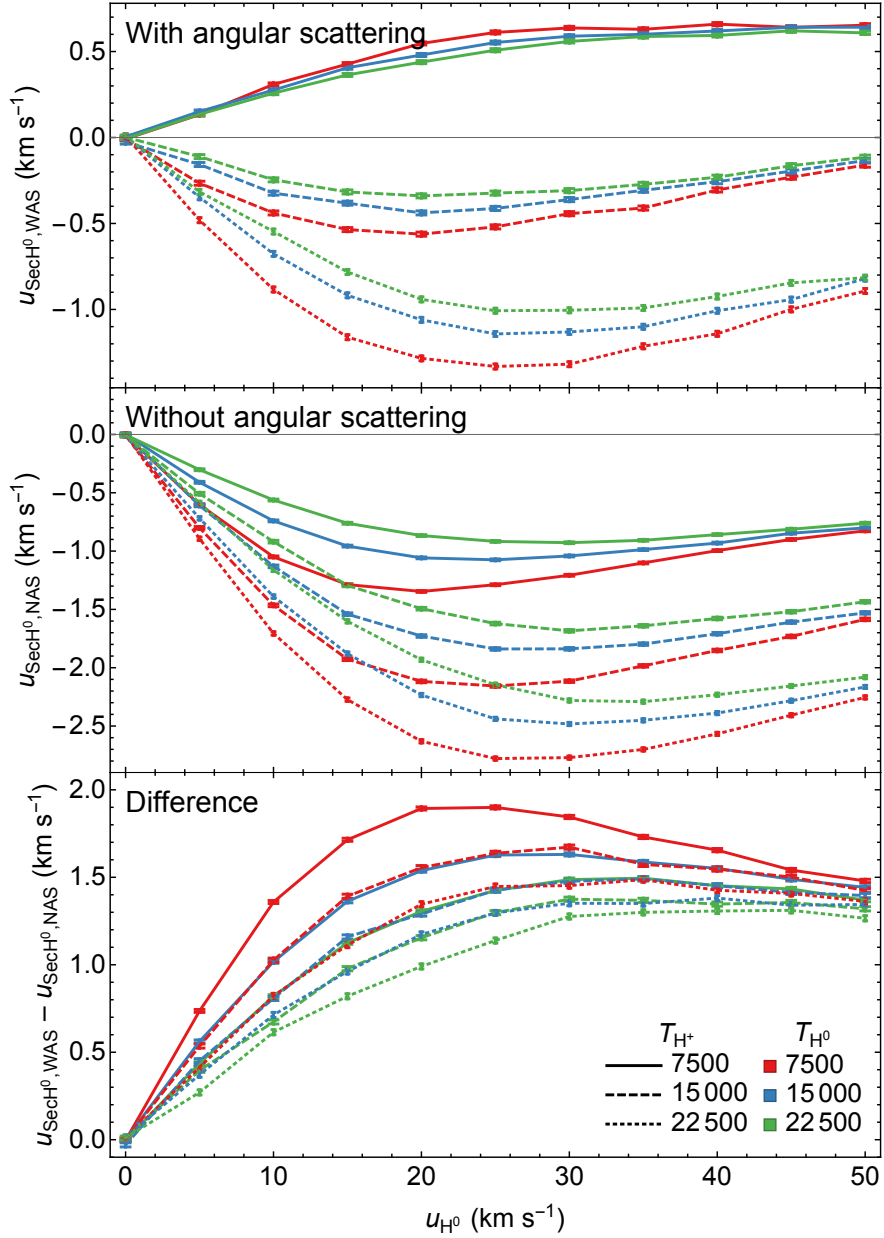

**Figure 5.** Bulk velocities of produced secondary hydrogen atoms in the plasma frame with (*top panel*) and without (*middle panel*) angular scattering and their difference (*bottom panel*) as a function of the bulk velocity of hydrogen atoms. Colors show three select temperatures of hydrogen atoms 7500 K (red), 15000 K (blue), and 22500 K (green), and the line style temperatures of protons 7500 K (solid), 15000 K (dashed), and 22500 K (dotted).

Accounting for angular scattering causes a partial momentum transfer to newly neutralized atoms from the parent primary atoms. For low temperatures of protons, the secondary atoms move in the same direction as the parent neutral atoms, i.e., oppositely to the case without angular scattering. However, for warm protons, the newly created secondary atoms move in the opposite direction to the parent neutrals, just like in the case without scattering. Interestingly, the temperature of the parent atoms seems to have much smaller impact. The difference between cases with and without angular scattering shows that this scattering transfer part of the momentum of neutrals atoms to the newly created secondary atoms. The net effect is the strongest for velocities of neutral atoms of ~20–30 km s$^{-1}$, where it can exceed 1.5 km s$^{-1}$. It is worth noting that the differences (as shown by the overlapping lines in the lower panel of Figure 5) are within numerical uncertainties of the integration for the cases with the same mean temperatures of protons and atoms, and the difference decreases as this sum increases.

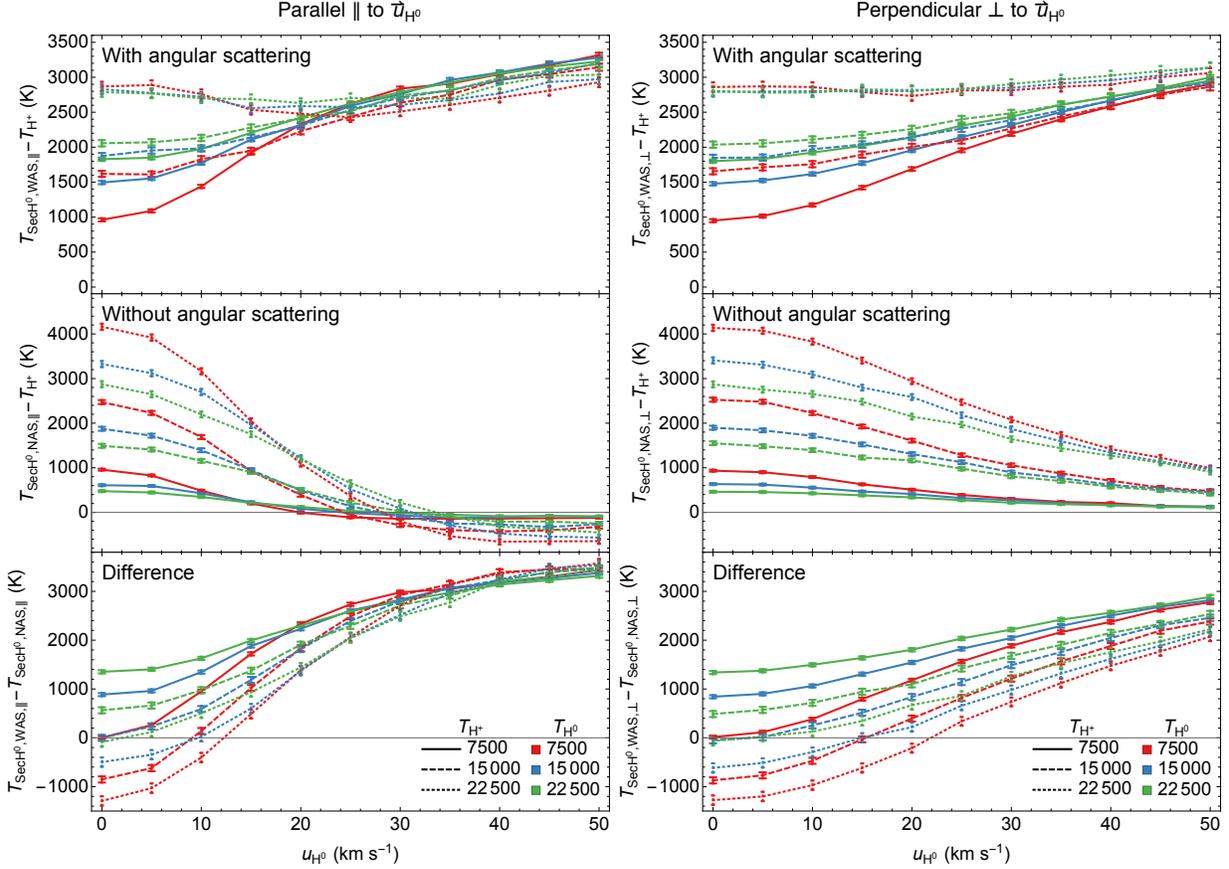

**Figure 6.** Parallel (*left panels*) and perpendicular (*right panels*) temperature components with respect to the relative bulk velocity of produced secondary hydrogen atoms with the temperature of the parent protons subtracted. The panels from top to bottom show the results with angular scattering, without angular scattering and their difference. Colors and line styles are the same as in Figure 5.

The temperature of secondary hydrogen atoms as a function of the relative bulk velocity of the parent populations for select temperatures of the parent populations is presented in Figure 6. Due to symmetry of the problem, we separate the parallel and perpendicular temperature components with respect to the relative bulk velocity. Without angular scattering, an increase in the temperature of secondary atoms is the highest for low relative motion of the parent populations. This increase is a result of higher charge exchange rate for protons with high thermal speeds and is therefore stronger for higher temperatures of protons. However, the effect becomes smaller as the temperature of neutral atoms increases. In this approach, the heating significantly drops for higher relative velocities of the hydrogen atoms. Dependence with the bulk velocity of neutral hydrogen is opposite when angular scattering is included in calculations because the scattering partially randomizes velocities of colliding particles. Differences of temperatures show that the scattering can increase the temperature of secondary neutral hydrogen atoms by up to ~3000 K compared to the approach without this scattering. Moreover, dependence on temperatures of the parent populations is smaller.

Asymmetries of secondary atom distributions given by the ratio of the parallel-to-perpendicular components of temperature are quite different in these two approaches, as shown in Figure 7. The magnitude of the asymmetry is similar in both situations; however, the parallel temperature is higher when angular scattering is included in the analysis, and the perpendicular component is higher for the approach without the scattering. Moreover, a hot neutral population tends to symmetrize the distribution, but the higher thermal energy of protons enhances the perpendicular component. This asymmetry of the distribution does not provide full information about the distribution of the secondary interstellar atoms. Discrepancies from the Maxwellian distribution for parallel and perpendicular components are discussed in the next section.

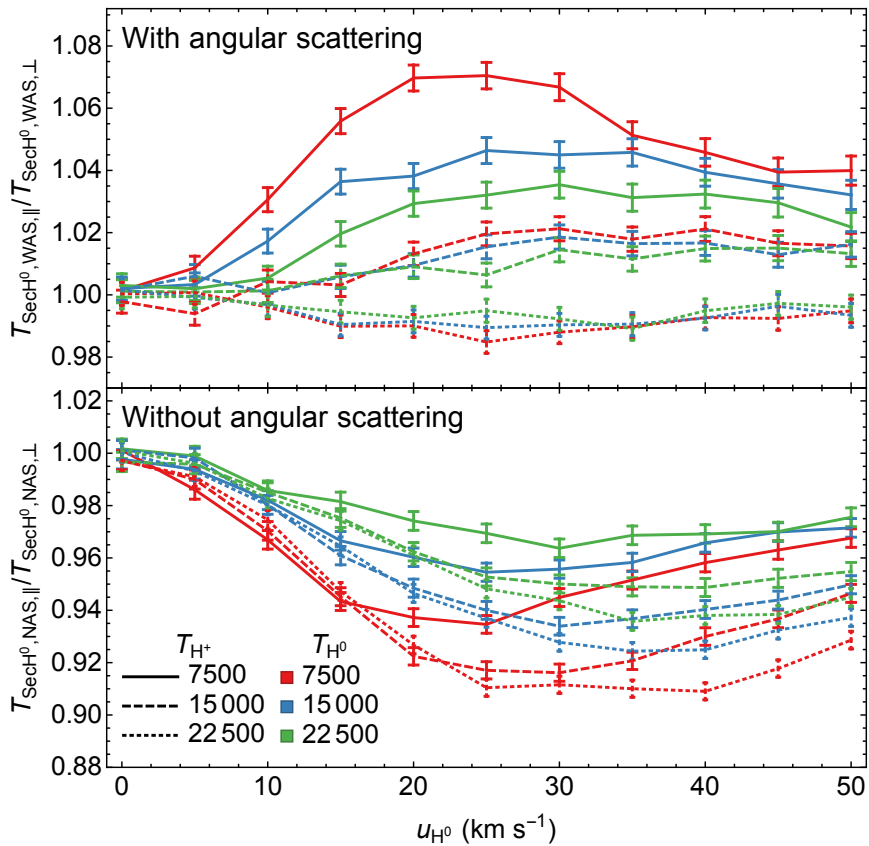

**Figure 7.** Temperature asymmetry of secondary hydrogen atoms. The ratio of the parallel to perpendicular temperatures is presented for the calculations with (*top panel*) and without (*bottom panel*) angular scattering. Colors and line styles are the same as in Figure 5.

*4.2 Distribution function of secondary hydrogen atoms*
The distribution functions of secondary atoms are obtained in this study using the Monte-Carlo integration (as discussed in Appendix B). In Section 4.1, we characterize these distributions with bulk speeds and temperatures that can provide Maxwell-Boltzmann approximations of the local production of secondary atoms. Compliance of these numerically obtained distributions with the Maxwell-Boltzmann distributions is checked here using the Kolmogorov-Smirnov statistics (e.g., DeGroot & Schervish 2012):

$$D_{K-S} = \sup|F(x) - F_{MB}(x)|, \tag{9}$$

where $F(x)$ is the empirical cumulative distribution function obtained from the Monte-Carlo integration, and $F_{MB}(x)$ is the cumulative Maxwell-Boltzmann distribution function with the parameters as determined in Section 4.1. We obtained the statistics separately for each cartesian component of the velocities of secondary atoms. Figure 8 presents the results for the parallel and the average of the two perpendicular components.

The Kolmogorov-Smirnov statistic can be used as a test of compliance of a random sample with a given distribution function. Based on the sample size of $10^6$, the value given by Equation (9) should not exceed 0.00136 at confidence level $\alpha = 0.05$. Therefore, the figure shows that the resulting distributions are not statistically consistent with the Maxwell-Boltzmann distribution. A marginal agreement is possible only without the scattering at a high relative speed of the parent populations. The discrepancy is the largest for the parallel component of the distribution with the scattering included in the analysis. In this case, the Kolmogorov-Smirnov statistic increases for high relative speeds of the parent population but is mitigated by the temperature of the parent protons.

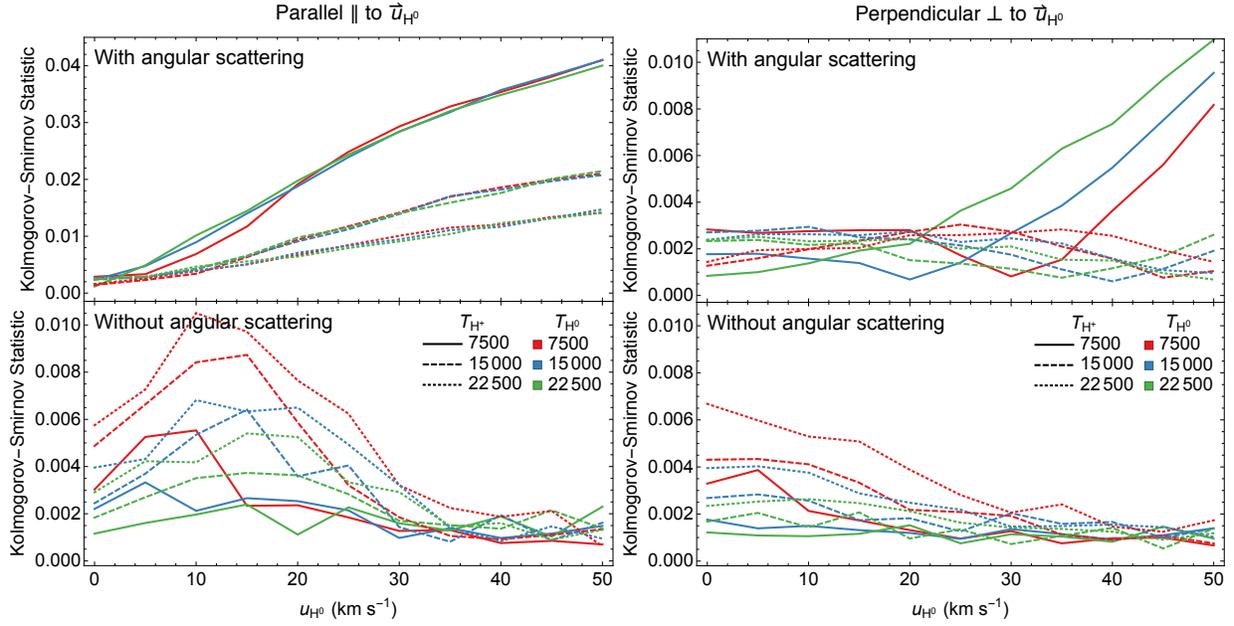

**Figure 8**. Kolmogorov-Smirnov statistics for parallel (*left panel*) and perpendicular (*right panel*) components of the distribution function for the calculations with (*top panel*) and without (*bottom panel*) angular scattering. Colors and line styles are the same as in Figure 5. Note that the scale in the top left panel is different than in the other panels.

Qualitative comparisons of histograms of velocities of secondary atoms with the Maxwell-Boltzmann distribution functions are presented in Figures 9 and 10. Figure 9 shows the results for the relative speed of the parent populations of 15 km s$^{-1}$, the temperature of protons of 15000 K, and temperature of hydrogen atoms of 7500 K. This represents values typical for the upwind part of the outer heliosheath downstream of the bow wave (Zank et al. 2013; Kubiak et al. 2019). For these conditions, the Maxwell-Boltzmann distributions well represent the parallel and perpendicular components of the secondary atom distributions. Figure 10 shows the results for the relative velocity of the parent populations of 45 km s$^{-1}$ and the temperatures of both populations of 7500 K. These conditions do not represent typical plasma in the outer heliosheath; however, they provide an illustration of significant differences between the considered approaches. The Maxwell-Boltzmann distribution well represents the results without angular scattering. Nevertheless, scattering introduces substantial asymmetry in the parallel component. For negative velocities, the distribution approximately follows the Maxwell-Boltzmann distribution with the parameters obtained from the case without momentum transfer, while for positive velocities, there is an extended tail that significantly exceeds the Maxwell-Boltzmann distribution. Moreover, for the perpendicular component, the tails of the distribution are slightly enhanced over the Maxwell-Boltzmann distribution.

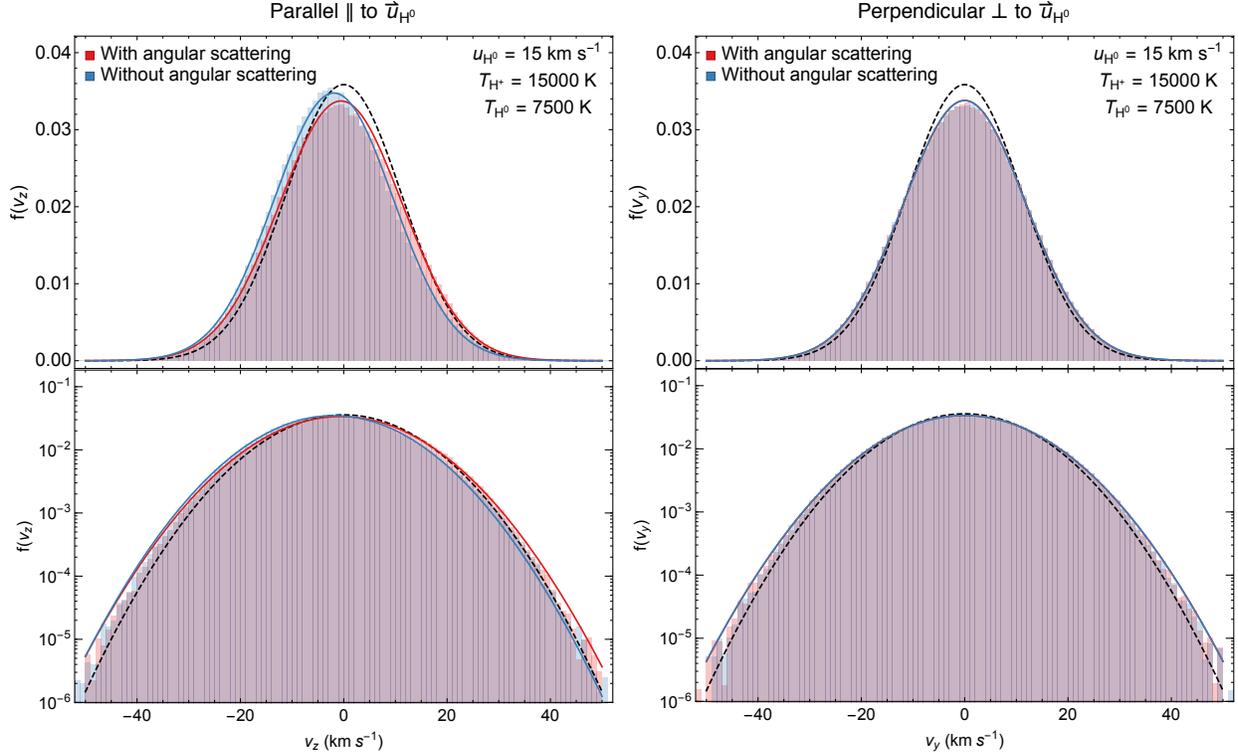

**Figure 9**. Distribution of velocities obtained from the Monte Carlo integration with (red histogram) and without (blue histogram) angular scattering in charge exchange collisions, obtained for the relative speed of the parent populations of 15 km s$^{-1}$, and temperatures $T_{H^+}$ = 15000 K and $T_{H^0}$ = 7500 K. The red and blue lines show the Maxwellian approximations with the parameters as presented in Figure 5 and 6. The gray dashed lines show the distribution of parent protons. The left and right panels show the result for the parallel and perpendicular components of the velocity, respectively. The top and bottom panels show the histograms in linear and logarithmic scales, respectively. The shift in the peak positions of the distributions shows the change of the mean velocity of the secondary hydrogen atoms. A small increase in the temperature for the parallel component is visible.

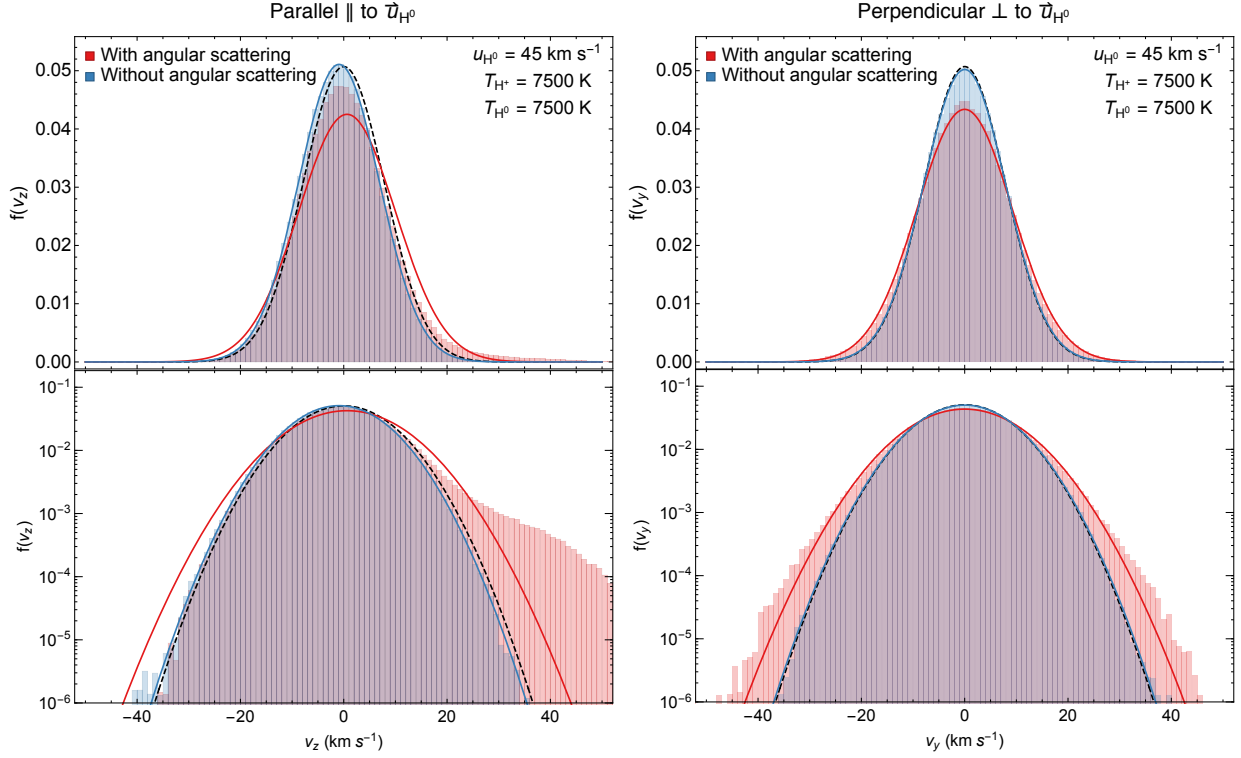

**Figure 10**. As in Figure 9, but for the relative speed of the parent populations of 45 km s$^{-1}$, and temperatures $T_{H^+} = 7500$ K and $T_{H^0} = 7500$ K. The high increase in the temperature of the secondary population (see also Figure 6) results from a one-side extended tail and is poorly represented by the Maxwell approximation.

## 5. Discussion

ISN atoms play the primary role in charge exchange collisions in the heliosphere and bring information about the physical state of the VLISM to the detectors located inside the heliosphere. Therefore, good knowledge of secondary atoms production is essential for many heliospheric studies. In this section, we briefly discuss some aspects of these studies that may be impacted by angular scattering in the charge exchange collisions outside the heliopause.

*5.1. Secondary ISN Helium Atoms*

The efficiency of the secondary ISN helium production is smaller due to lower charge exchange rate resulting from an order of magnitude smaller density of He$^+$ ions compared to protons in the VLISM. As a result, the secondary population comprises only a small portion of all ISN helium atoms, and can be separated from the primary population (Kubiak et al. 2014, 2016; Wood et al. 2017). Bzowski et al. (2017, 2019) analyzed the secondary population of ISN helium atoms jointly with the primary population by synthesizing the neutral atom signal in the outer heliosheath using a global model of the heliosphere (Heerikhuisen & Pogorelov 2011; Heerikhuisen et al. 2015; Zirnstein et al. 2016). In their study, secondary atom production is considered without angular scattering. They found that the synthesized model is better compared to the earlier model with two Maxwell-Boltzmann components; however, they found that their reduced $\chi^2$ measure is still significantly too high, suggesting missing physical effects in their analysis.

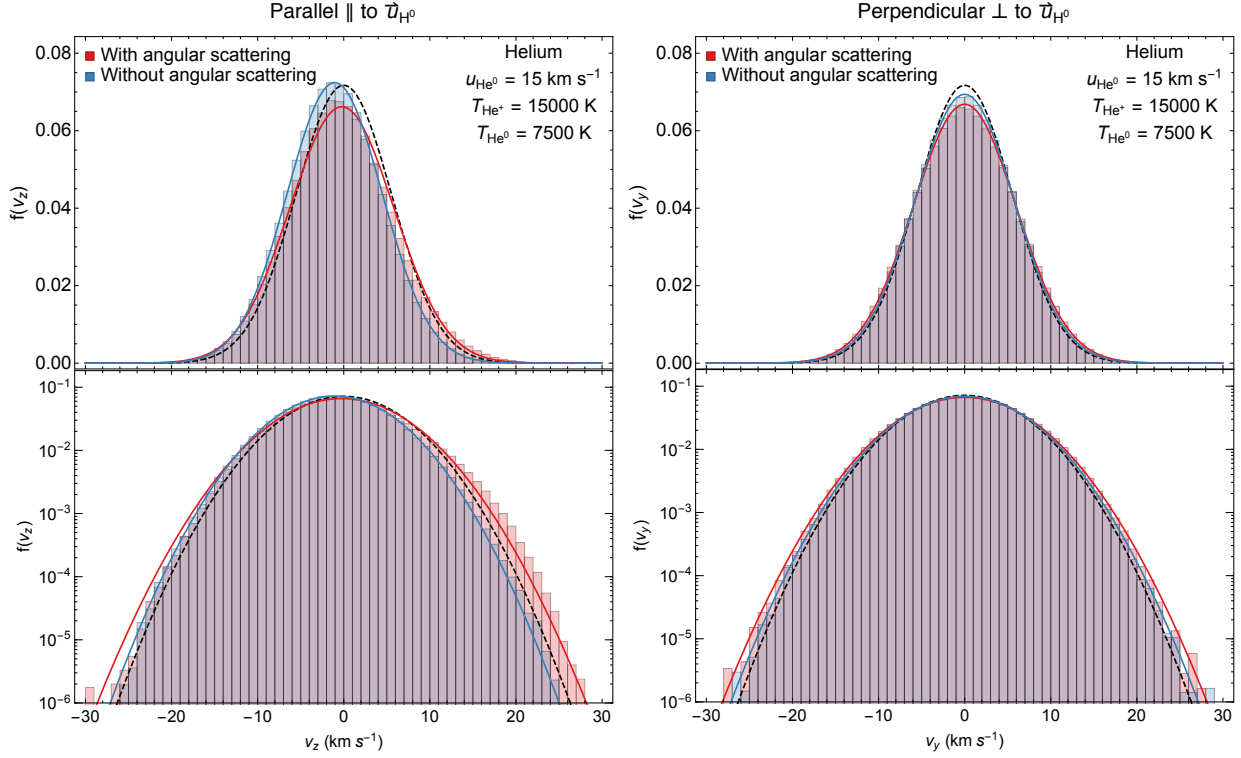

**Figure 11**. As in Figure 9 but for helium atoms assuming the differential cross section scaling (see text).

In this paper, we focus on ISN hydrogen atoms, but a simple estimation of the momentum transfer effect can be obtained assuming the same shape of the differential cross section in $He^0 – He^+$ charge exchange collisions as for proton – hydrogen atom collisions for the same energy per mass (Lewkow et al. 2012; Lewkow 2016). Based on this estimation, the distribution of secondary ISN helium atoms is presented in Figure 11 for the relative parent population velocity of 15 km s$^{-1}$, and temperatures of He$^+$ ions and He atoms of 15000 K and 7500 K, respectively. The parameters correspond to those used to plot Figure 9 for hydrogen atoms. For helium, the effect on these parameters is significantly higher, and deviation from the Maxwell distribution is clear. Accounting for angular scattering change the bulk velocity by 0.8 km s$^{-1}$ and the parallel and perpendicular temperature by ~2900 K and ~1200 K, respectively. Helium ions and atoms of the same temperature as protons and hydrogen atoms have approximately half the thermal velocity, so a significant departure from the thermalized population is expected to start at lower relative velocities of the parent populations.

*5.2 Density Distribution of Hydrogen Atoms*

The charge exchange process between protons and hydrogen atoms significantly modifies the distribution of ISN hydrogen atoms in the heliosheath. As a result, the density of the ISN hydrogen inside the heliopause is smaller compared to the density in the VLISM (e.g., Müller et al. 2008). In the outer heliosheath, the density of the ISN hydrogen is significantly increased, forming the hydrogen wall (Baranov & Malama 1993). In the heliosphere, ISN hydrogen atoms are ionized mostly by charge exchange with solar wind protons, so their density is significantly depleted at 1 au. At the same time, the ionized ISN atoms are a source of pickup ions (PUIs), which form an energetic population in the solar wind (McComas et al. 2017) and decelerate the expanding solar wind (Richardson et al. 2008; Elliott et al. 2019). PUIs also play an important role in the acceleration processes at interplanetary shocks and at the termination shock (Zank et al. 2014, 2018; Zirnstein et al. 2018).

Angular scattering in charge exchange collisions in the outer heliosheath does not impact the production rate but instead changes the distribution of secondary atoms. They are warmer and deflected in the direction of the motion of parent neutrals in the plasma frame (Section 4.1). In the outer heliosheath, the secondary atoms are preferentially directed inside the heliosphere compared to the case without this scattering. Therefore, the density of the ISN hydrogen can be increased inside the heliopause but decreased in the outer heliosheath, including in the hydrogen wall.

Here, we build a simple "toy" model to show the effect of the momentum transfer on the distribution of the ISN hydrogen. For this purpose, we solve the following one-dimensional differential equation:

$$v \frac{\partial f_\text{H}(r,v)}{\partial r} = -n_\text{p}(r) v_\text{rel} \sigma_\text{cx}(v_\text{rel}) f(r,v) + \frac{\partial f_\text{H}(r,v)}{\partial t}\bigg|_\text{P} \quad (10)$$

where $n_\text{p}$ represents the proton density, $r$ is the distance from the Sun, $v_\text{rel}$ is the mean relative velocity between an atom moving with velocity $v$ and the Maxwell-Boltzmann distribution of protons. The last term represents the production rate calculated with and without momentum transfer. We solve this equation using plasma density, flow, and temperature from the MHD model (Heerikhuisen et al. 2015) with the parameters found by Zirnstein et al. (2016) along the inflow direction of the VLISM. Both terms were used outside the heliosphere. Inside the heliosphere, we only use a loss term approximated as $\beta_0 r^{-2} f(r,v)$ with $r$ expressed in au, and $\beta_0 = 6 \times 10^{-7}$ s$^{-1}$ for the total ionization rate at 1 au (Sokół et al. 2019). We solve Equation (10) with $x$ representing the distance from the Sun, and therefore negative and positive velocities $v$ correspond to atoms flowing toward and away from, respectively, the Sun. We assume ISN hydrogen atoms are represented by a Maxwell-Boltzmann distribution at the outer boundary of the model (at ~1000 au).

Results of the model are shown in Figure 12. Inclusion of the momentum transfer in the model enhances the density of atoms in the heliosphere by 4–7%, with the most significant enhancement closest to the Sun. This figure also shows that the distribution of ISN hydrogen atoms is slightly enhanced at higher velocities relative to the Sun. Atoms with higher velocities can be more efficiently detected at 1 au. Therefore, the effect is an important factor necessary to understand observations of ISN hydrogen atoms from IBEX-Lo (Schwadron et al. 2013; Katushkina et al. 2015) and, in the future, also from IMAP-Lo (McComas et al. 2018). Detailed calculations of the transport in full three-dimensional model are, however, beyond the scope of this study.

*5.3 Momentum Transfer to Plasma*

Charge exchange between ions and neutrals in the outer heliosheath play another essential role for the global structure of the heliosphere (Pogorelov et al. 2007, 2017). This process changes the momenta of plasma and neutrals because it replaces protons with the kinetic properties of the plasma with new ones that inherit kinetic properties from ISN atoms and, as a result, tends to reduce asymmetries in the global heliosphere. Angular scattering in charge exchange collisions as discussed in this analysis may, therefore, impact this process. However, significance of momentum transfer increases with collision energy because the charge exchange rate increases with energy and larger momenta are transferred in high energy collisions. Hence, contribution of lower energy collisions in momentum transfers is small regardless of the angular scattering.

Heerikhuisen et al. (2009) noted that even the isotropic scattering assumed in the charge exchange process does not significantly impact the global structures of the heliosphere, and conclude that this effect can be neglected from the perspective of the global modeling.

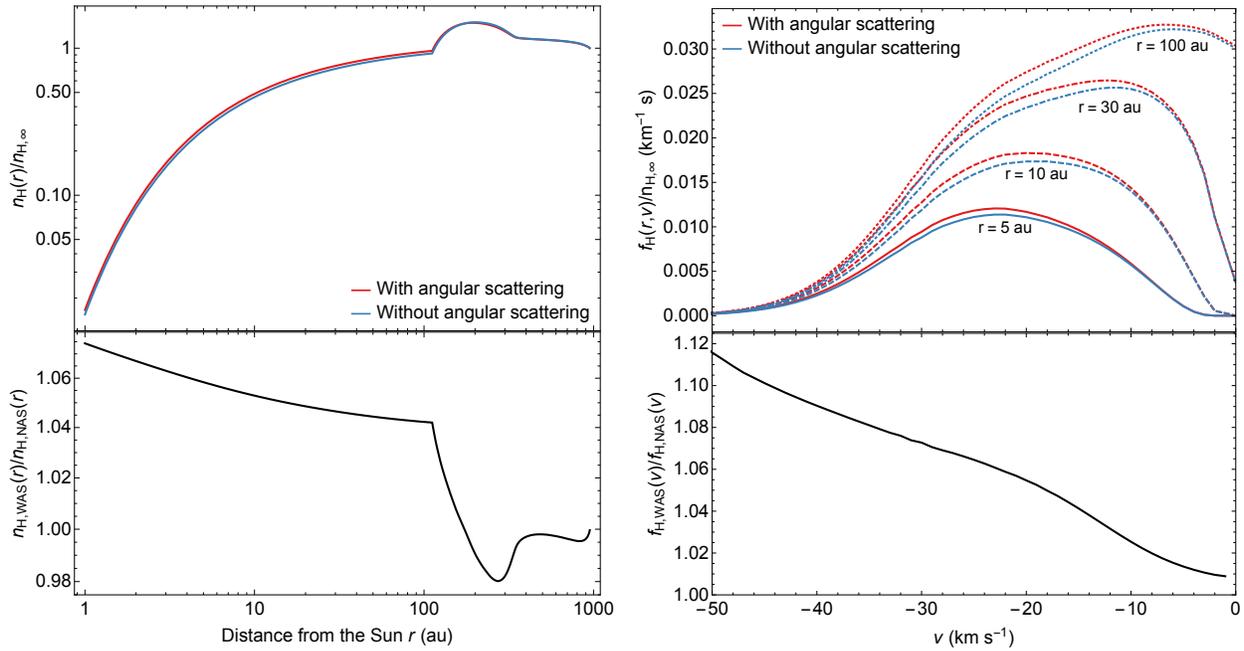

**Figure 12.** *Left panel*: Density of ISN hydrogen compared to the density in the VLISM in a one-dimensional model (see text) with and without angular scattering in charge exchange collisions as a function of distance from the Sun. The bottom panel shows the ratio of these two cases. *Right panel*: Distribution functions of ISN hydrogen atoms in the heliosphere in the one-dimensional model for the select distances from the Sun. The ratio (bottom part) between cases with and without scattering is higher for faster atoms (relative to the Sun) but it is the same at all distances from the Sun inside the heliosphere.

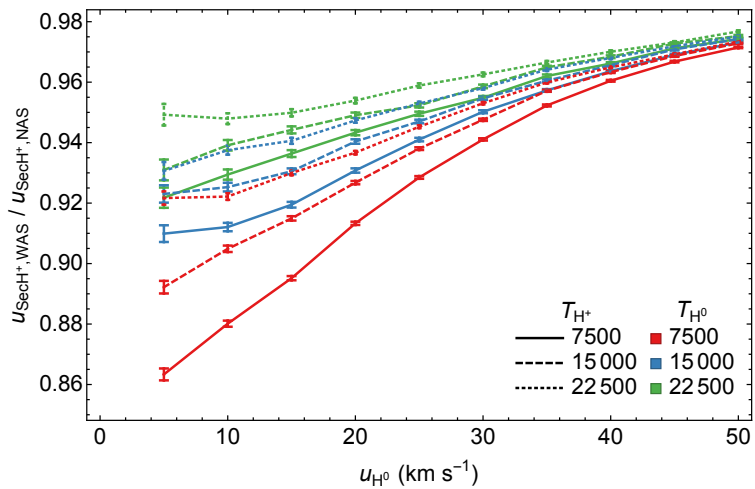

**Figure 13.** Ratio of the bulk velocities of newly created protons in the plasma frame in the case with and without angular scattering in charge exchange collisions. The lines and colors are as in Figure 5.

The importance of momentum transfer to the plasma can be studied in the two considered cases based on comparison of the mean velocity of secondary protons, i.e., created from ISN hydrogen atoms as a result of charge exchange collisions in these two situations. Note that we discuss here the mean velocity of protons just after the charge exchange, i.e., this represents their instantaneous velocities at which they are picked up by the magnetic field. The ratio of these velocities is presented in Figure 13. The largest effect is expected for the low relative difference in the bulk velocities of the parent distributions. However, under such conditions, the transferred momentum is small. In all analyzed cases, the ratio is relatively high in the range of 0.86–0.95 for the relative velocities of parent populations below 25 km s$^{-1}$. This result explains why angular scattering in the charge exchange process plays only a secondary role in the shaping of the global structure of the heliosphere (Heerikhuisen et al. 2009).

*5.4 Elastic Scattering*

Angular scattering is also present in elastic collisions, in which interacting particles do not exchange their electrical charge but only part of their momentum. A simple application of these collisions in a global model of the heliosphere was presented by Williams et al. (1997). They analyzed the role of elastic collisions of hydrogen atoms on the properties of ISN hydrogen atoms in the heliosphere. These authors noticed that there is a significant difference inside the heliosphere, where the elastic collisions lead to fast thermalization of the different populations of hydrogen atoms. Izmodenov et al. (2000) argued that the role of the elastic scattering was significantly overestimated by Williams et al. (1997).

Recently, Gruntman (2013, 2018) analyzed the impact of elastic scatterings between ISN helium atoms and solar wind ions. Their calculations show an apparent increase by ~175–270 K in the temperature observed by IBEX or Ulysses compared to the temperature of the VLISM. Additionally, a halo of atoms visible as enhanced wings in the distribution can be produced by this process. Another study performed by Kubiak et al. (2014 Appendix B) estimated a thermalization distance for two components of ISN helium. This study showed that the distance necessary to fully thermalize two populations tens of thousands of au. However, significant deviation in the kinetic properties of the original population can be expected after only about a thousand au.

**6. Summary**

Charge exchange between protons and hydrogen atoms plays an important role in all heliospheric studies. However, these collisions are commonly assumed to preserve momenta of the colliding particles, so that the daughter particles conserve velocities of the parent particles, i.e., outgoing hydrogen atoms and protons have the same velocity as incoming protons and hydrogen atoms, respectively. This assumption is equivalent to requiring no scattering in these collisions. In this study, we estimated the impact of this scattering on the properties of the newly created secondary ISN hydrogen atoms.

We modeled collisions of the parent Maxwell-Boltzmann populations with temperatures from 7500 to 22500 K, and relative velocities less than 50 km s$^{-1}$. We performed Monte Carlo integration of the collision terms with and without momentum transfer due to angular scattering in charge exchange collisions. We found that the momentum transfer leads to the secondary population of ISN hydrogen atoms with velocities increased in the direction of motion of the primary population of ISN hydrogen atoms by up to ~2 km s$^{-1}$, with a maximum at relative velocity of the parent population in the range ~15–25 km s$^{-1}$ (Section 4.1). Additionally, the resulting population is heated by up to ~3000 K and shows a small asymmetry in the directions parallel and perpendicular to the relative velocity of the parent population (Section 4.1).

Moreover, for the high relative velocity of the parent populations, the produced secondary population significantly departs from the Maxwell-Boltzmann distribution (Section 4.2), especially in the direction parallel to the relative motion of the parent populations.

In Section 5.1, we showed that angular scattering in charge exchange might also be important for the production of the secondary population of ISN helium atoms, which are generally less processed in the heliosphere and thus can give better insight into this process. This scattering also leads to change in a distribution of the ISN hydrogen in the heliosphere, with increased density in the heliosphere and decreased in the region of the hydrogen wall (Section 5.2). On the other hand, this process does not significantly affect the momentum transfer from ISN hydrogen atoms to plasma in the outer heliosheath (Section 5.3). We note that a similar effect may be related to elastic collisions that are also mostly neglected in the heliospheric studies (Section 5.4).

The obtained results suggest that angular scattering should be accounted for in analyses of the secondary populations of ISN atoms observed by detectors at 1 au. Increased sensitivities of future detectors as expected on *Interstellar Mapping and Acceleration Probe* (*IMAP*, McComas et al. 2018) can give a better insight into the role of angular scattering in binary collisions. Moreover, this process may be necessary to fully understand observations of PUIs, especially in the outer heliosphere, as currently possible using the SWAP instrument on New Horizons (McComas et al. 2008, 2017).

*Acknowledgments*: This work was funded by the *IBEX* mission as part of NASA's Explorer Program (80NSSC18K0237) and IMAP mission as a part of NASA's Solar Terrestrial Probes (STP) Program (80GSFC19C0027). J. H. and E. J. Z. acknowledge support from NASA grant 80NSSC18K1212.

**Appendix A. Rotations of scattered velocity vectors**

Scattering in the CM reference frame $\boldsymbol{r}(\boldsymbol{v}, \theta, \phi)$ is a superposition of two rotations: first by angle $\theta$ around a vector perpendicular to vector $\boldsymbol{v}$, and the result is rotated around vector $\boldsymbol{v}$ by angle $\phi$. In general rotation of vector $\boldsymbol{r}$ by angle $\alpha$ around normalized vector $\boldsymbol{n}$ can be written as (Koks 2006, Chapter 4)

$$\boldsymbol{R}_{\boldsymbol{n}}(\boldsymbol{r}, \alpha) = (1 - \cos \alpha)(\boldsymbol{n} \cdot \boldsymbol{r})\boldsymbol{n} + \cos \alpha \, \boldsymbol{r} + \sin \alpha \, \boldsymbol{n} \times \boldsymbol{r}. \tag{11}$$

For the first rotation, we need to select a vector perpendicular to the original vector. This selection is arbitrary and correspond to symmetry in rotation by angle $\phi$. We chose the first rotation axis as $\boldsymbol{a} = \boldsymbol{v} \times [0,0,1]$. Hence, the sought rotation is given by the following superposition

$$\boldsymbol{r}(\boldsymbol{v}, \theta, \phi) = R_{\boldsymbol{v}/|\boldsymbol{v}|}\big(R_{\boldsymbol{a}/|\boldsymbol{a}|}(\boldsymbol{v}, \theta), \phi\big). \tag{12}$$

This formula calculated with components of vector $\boldsymbol{v}$ provides

$$r\left(\begin{bmatrix}v_x\\v_y\\v_z\end{bmatrix},\theta,\phi\right)=\begin{bmatrix}v_x\cos\theta+\dfrac{\left(-v_xv_z\cos\phi+v_y\sqrt{v_x^2+v_y^2+v_z^2}\sin\phi\right)\sin\theta}{\sqrt{v_x^2+v_y^2}}\\v_y\cos\theta+\dfrac{\left(-v_yv_z\cos\phi-v_x\sqrt{v_x^2+v_y^2+v_z^2}\sin\phi\right)\sin\theta}{\sqrt{v_x^2+v_y^2}}\\v_z\cos\theta+\sqrt{v_x^2+v_y^2}\cos\phi\sin\theta\end{bmatrix}. \qquad (13)$$

**Appendix B. Monte Carlo Integration of Collision Term**

In this study, we use the Monte Carlo method to calculate the integral given in Equation (6). Here, we briefly present the adopted scheme. We performed calculations with $i_{\max}=10^6$ collisions for select temperatures of protons and hydrogen atoms.

1. Pairs $\{v_{p,i},v_{H,i}\}_{i=1,\ldots,i_{\max}}$ of the initial velocities of the incoming protons $v_{p,i}$ and hydrogen atoms $v_{H,i}$ are drawn from the Maxwell distributions assumed for the parent particles. Note that the Maxwell distribution for proton is assumed at zero bulk velocity, whereas for hydrogen atoms, the bulk velocity is along the Z-axis.
2. For each pair, the CM energy $E_{CM,i}$ is calculated from the CM velocity, see Equation (8).
3. The scattering angle $\theta_i$ is drawn from a normalized differential cross section for the nearest logarithmically spaced energy bin from those given by Schultz et al. (2016). The azimuthal angle $\phi_i$ is drawn from the range $[0,2\pi]$ using a uniform distribution.
4. Velocities of secondary atoms $v_{secH,i}$ are calculated using Equations (7-8, 13).
5. For each collision, a weight given as $w_i=|v_{p,i}-v_{H,i}|\sigma_{cx}(|v_{p,i}-v_{H,i}|)$ is calculated that provides the probability of the charge exchange pair.

The distribution of the secondary atoms is evaluated from the set of the velocities of the secondary atoms with the appropriate weights $\{v_{secH,i},w_i\}_{i=1,\ldots,i_{\max}}$. For this study, we calculate the properties of the resulting distribution using this procedure. The parameters are known with a finite precision due to the random character of this procedure.